\documentclass[twocolumn,showpacs,superscriptaddress,english,aps,prl]{revtex4}
\usepackage[T1]{fontenc}
\usepackage[latin9]{inputenc}
\usepackage{amstext}
\usepackage{graphicx}
\usepackage{amssymb}
\usepackage{bm}

\makeatletter

\providecommand{\LyX}{L\kern-.1667em\lower.25em\hbox{Y}\kern-.125emX\@}
\newcommand{\lyxmathsym}[1]{\ifmmode\begingroup\def\b@ld{bold}
  \text{\ifx\math@version\b@ld\bfseries\fi#1}\endgroup\else#1\fi}

\@ifundefined{textcolor}{}
{%
 \definecolor{BLACK}{gray}{0}
 \definecolor{WHITE}{gray}{1}
 \definecolor{RED}{rgb}{1,0,0}
 \definecolor{GREEN}{rgb}{0,1,0}
 \definecolor{BLUE}{rgb}{0,0,1}
 \definecolor{CYAN}{cmyk}{1,0,0,0}
 \definecolor{MAGENTA}{cmyk}{0,1,0,0}
 \definecolor{YELLOW}{cmyk}{0,0,1,0}
 }

\makeatother

\usepackage{babel}

\begin{document}

\title{The experimental observation of quantum Hall effect of $l = 3$ chiral charge carriers in trilayer graphene.}

\author{Liyuan Zhang}
\affiliation{CMPMSD, 
Brookhaven National Laboratory, Upton, NY 11973 USA}
\affiliation{Department of Physics, Renmin University of China, Beijing, PRC}

\author{Yan Zhang}
\affiliation{Department of Physics and Astronomy, Stony Brook University, Stony Brook, New York 11794-3800, USA}

\author{J. Camacho}
\affiliation{CMPMSD, 
Brookhaven National Laboratory, Upton, NY 11973 USA}
\affiliation{Nanotechnology Center for Collaborative R$\&$D, University of Wisconsin-Platteville, WI  53818 USA}

\author{M.~Khodas}
\affiliation{CMPMSD, 
Brookhaven National Laboratory, Upton, NY 11973 USA}
\affiliation{Department of Physics and Astronomy, University of Iowa, Iowa City, IA 52242 USA}

\author{I.~A.~Zaliznyak}
\email{zaliznyak@bnl.gov}
\affiliation{CMPMSD, 
 Brookhaven National Laboratory, Upton, NY 11973 USA}

\begin{abstract}

Low-energy electronic states in monolayer and bilayer graphenes present chiral charge carriers with unique and unusual properties of interest for electronic applications \cite{Novoselov2005,Zhang2005,Novoselov2006}. Here, we report the magnetotransport measurements in the ABC-stacked trilayer graphene as a function of charge carrier density, magnetic field, and temperature, which show clear evidence of $l = 3$ chiral quasiparticles with cubic dispersion \cite{CastroNeto2009,Guinea2006,Aoki2007,Ezawa2007,KoshinoAndo2007,KoshinoMcCann2009,Koshino2010,ZhangMacDonald2010}, existing in a large, $~20×60~\mu$m$^2$ device. Shubnikov-deHaas oscillations (SdHO) reveal the Berry's phase $3\pi$, and the marked increase of cyclotron mass near charge neutrality, consistent with divergent behavior expected for $l=3$ quasiparticles. We also observe the predicted unconventional sequence of quantum Hall effect (QHE) plateaus, $\sigma_{xy} = ±6e^2/h,  ±10e^2/h, ...$.

\end{abstract}


\maketitle



The discovery of graphene presented not only realizations of truly two-dimensional (2D) electronic systems, but revealed entirely new, exotic electronic states.  In monolayer graphene the low-energy electronic structure is described by chiral quasiparticles with linear dispersion, they obey the relativistic Dirac equation and have a Berry phase of $\pi$ \cite{Novoselov2005,Zhang2005}. In bilayer graphene \cite{Novoselov2006}, the $l = 2$ chiral charge carriers have quadratic dispersion and Berry's phase $2\pi$. Unusual properties are revealed in magnetotransport experiments in a perpendicular magnetic field, where the Berry phase determines the shift of Shubnikov-deHaas resistance oscillations, while their period is governed by filling of Landau levels (LLs) of chiral 2D quasiparticles. The quantum Hall effect shows an unconventional sequence of plateaus of Hall conductivity, $\sigma_{xy}$, with quantized steps of $4e^2/h$, except for the first plateau, where it is governed by the Berry's phase.

Despite significant interest in studying layered graphene systems with more than two layers, experimental progress has been limited \cite{Craciun2009,Zhu2009,Liu2010,Bao2010,Mak2010}. Low-energy electronic properties depend crucially on the stacking order of graphene layers \cite{CastroNeto2009,Guinea2006,Aoki2007,Ezawa2007,KoshinoAndo2007,KoshinoMcCann2009,Koshino2010,ZhangMacDonald2010,Mak2010}, and therefore such studies require samples with well-defined stacking sequence. In a bilayer, two honeycomb nets of carbon atoms are positioned with half of the atoms of the top layer (B) right above the atoms of the bottom layer (A) and the other half at the centers of the hexagonal voids in it. The third carbon net in a trilayer can either be placed with its atoms above the atoms of the bottom layer A, as in Bernal structure of crystalline graphite \cite{Friese1962}, or with its voids above the lined-up atoms pairs in layers A and B, thus breaking the reflection symmetry (Fig. 1a). The latter, ABC stacking, is found in metastable rhombohedral modification of graphite \cite{Friese1962}.

The electronic structure of graphene multi-layers is derived from the hybridization of monolayer states via interlayer hopping. Its main features are captured already by only considering hopping between the nearest-neighbour carbons, which are stacked above each other in two adjacent layers, $\gamma_1 \sim 0.1\gamma_0$, Fig. 1a ($\gamma_0 \approx 3.16$ eV is the intra-layer hopping, in bulk graphite $\gamma_1 \approx 0.4$ eV, and there are also further-neighbour hoppings, $\gamma_2 - \gamma_5$, which are not shown) \cite{KoshinoAndo2007,KoshinoMcCann2009,Koshino2010,ZhangMacDonald2010}. In bilayer, low-energy electronic states retain the chiral character but have flatter, quadratic dispersion. The effective Hamiltonian is the nonlinear generalization of Dirac-Weyl quasiparticles of the monolayer,
\begin{equation}
\label{Hamiltonian-l}%
\hat{H}_l = v_l\xi^l \left(
\begin{array}{cc} 0 & (\hat{\pi}^+)^l \\ \hat{\pi}^l & 0 \end{array} \right) = \frac{v_l \hat{p}^l \xi^l}{2} \left( \hat{\sigma}^- e^{i l \xi \hat{\varphi}_{\bf p}} + c. c. \right),
\end{equation}
with $l = 2$. Here, $v_l = v^l/\gamma_1^{l-1}$, where $v \approx 10^6$ m/s is the velocity of Dirac fermions in the monolayer, $(\hat{p}_x, \hat{p}_y) = \hat{p}(\cos \hat{\varphi}_{\bf p}, \sin \hat{\varphi}_{\bf p})$ is the 2D momentum, $\hat{\pi} = \hat{p}_x+i \hat{p}_y$, $\hat{\sigma}^{\pm} = \hat{\sigma}_x \pm i \hat{\sigma}_y$ are the pseudo-spin Pauli matrices, and $\xi = ±1$ is a valley index. There are two low-energy valleys with opposite chirality. For our purposes they can be considered non-interacting and only yielding a two-fold degeneracy of all states; together with the two-fold electron spin degeneracy this gives a factor of 4 in the Hall conductivity quantization, $4e^2/h$. The wave functions of such degree-$l$ chiral fermions acquire a Berry phase of $l\pi$ upon an adiabatic propagation along a closed orbit. In bilayer, this results in $l = 2$, $2\pi$ chiral quasiparticles and an unusual integer QHE sequence with a double step, $\sigma_{xy} = 8e^2/h$, between the hole and electron gases across the N=0 LL observed in experiment \cite{Novoselov2006}.

The low-energy band structure of ABA stacked graphene trilayer consists of superimposed linear and quadratic spectra. Hence, transport is governed by two types of chiral quasiparticles: monolayer-like massless ($l = 1$) and bilayer-like massive ($l = 2$), albeit with a larger effective mass than in bilayer, $m_{ABA} =  \sqrt{2}m_{AB} \approx  0.05m_e$ ($m_{AB} = \gamma_1/(2v^2) \approx 0.035m_e$, where $m_e$ is the electron mass) \cite{Guinea2006,KoshinoAndo2007}. In contrast to bilayer, where application of an electric field across layers opens a band gap in the spectrum \cite{Zhang2009}, in ABA trilayer it actually leads to a tuneable band overlap \cite{Craciun2009,Zhu2009,Liu2010,Bao2010,Mak2010}.

While to our knowledge no magnetotransport experiments on ABC trilayers have been published, this case is actually most interesting, as it is expected to present new, $l = 3$ chiral quasiparticles with cubic dispersion \cite{CastroNeto2009,Guinea2006,Aoki2007,Ezawa2007,KoshinoAndo2007,KoshinoMcCann2009,Koshino2010,ZhangMacDonald2010,low_n_note},  $\varepsilon(p) = \gamma_1(vp/\gamma_1)^3$, Fig. 1b. That this situation is quite remarkable could already be seen from the fact that the effective mass of such charge carriers is energy-dependent and diverges at the charge neutrality point (CNP), $m_{ABC} = p\partial p /\partial \varepsilon = \gamma_1^2/(3pv^3) = \frac{2}{3}m_{AB}(\gamma_1/\varepsilon)^{1/3}$. This corresponds to a diverging low-energy density of states, $D(\varepsilon) \sim \varepsilon^{-1/3}$, in contrast to a constant $D(\varepsilon)$ in bilayer and a vanishing one in monolayer. Such abundance of low-energy scattering states would make non-chiral fermions with cubic dispersion unstable with respect to decays -- but in ABC trilayer $l = 3$ quasiparticles are protected by chirality conservation. On the other hand, diverging $D(\varepsilon)$ leads to super-polarizability and super-efficient screening, which were predicted theoretically in perpendicular electric field \cite{Koshino2010}.

%
\begin{figure}[!t]
\vspace{-0.25in}
\includegraphics[width=1.\linewidth]{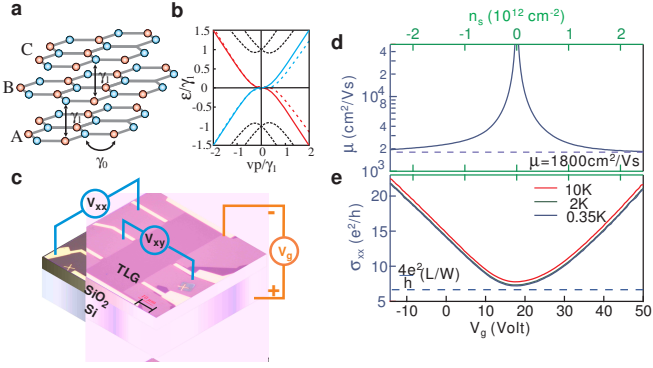}
\caption{a, The structure, and b, the low-energy band structure \cite{KoshinoMcCann2009}, of ABC-stacked TLG. c, The image of our large-area etched Hall bar and schematics of the measurement setup. The black scale bar is $10 \mu m$. d, Field mobility, $\mu = \sigma_{xx}/(n_se)$, at T = 0.35 K and e, longitudinal conductivity, as a function of BG voltage $V_g$, or carrier density induced via electric field effect, $n_s=C_gV_g$ (top scale). The minimum conductivity is slightly larger than $6e^2/h$, which agrees with the conductance via four quantum channels (dashed line), indicating a quantum-mechanical origin.
 }
\label{Fig1:mobility}
\vspace{-0.1in}
\end{figure}
%


Here we report measurements of magnetotransport oscillations and quantum Hall effect in the ABC-stacked trilayer graphene. The optical image of our Hall bar device and schematics of the measurement setup are shown in Fig. 1c. Our sample was prepared following standard procedure \cite{Novoselov2004} by mechanical exfoliation of kish graphite. While the number of layers can already be identified by optical contrast, it was also confirmed using Raman micro-spectroscopy, by measuring the full width at half maximum (FWHM) of the 2D Raman band \cite{Ferrari2006,Graf2007}. The characteristic asymmetric shape of the 2D-band spectrum, with dip near 2700 cm$^{-1}$, was used to identify the ABC stacking in our sample \cite{Hao2010,Lui2011}.


The transport properties of the ABC trilayer appear enigmatic: a na\"{i}ve estimate of the impurity-limited mobility following similar arguments to the monolayer case gives an energy (charge density) dependent mobility, $\mu \sim \varepsilon^{2/3} \sim n_s$. This clearly disagrees with our transport measurements of the ABC trilayer device of Fig. 1c shown in Fig. 1d,e. We find that the gate voltage dependence of longitudinal conductivity, $\sigma_{xx}$, is very similar to that of monolayer/bilayer graphene, with roughly linear asymptotic behavior at high charge carrier densities, $n_s$ (or gate voltages $V_g$), for both polarities ($+/-$ correspond to electrons/holes). The asymptotic field mobility estimated assuming a simple Drude model is $\mu =  \sigma_{xx}/(en_s) \approx 1800$ cm$^2$V$^{-1}$s$^{-1}$ ($e$ is an electron charge) in our trilayer, which is lower than in similar high quality monolayer samples. Nevertheless, the maximum resistance near the CNP ($V_g \approx 17.3$ V), $R_0 \approx 5.7$ k$\Omega$, roughly agrees with the quantum conductance via 4 degenerate channels, $h/(4e^2) \approx 6.5$ k$\Omega$, which is expected for quantum transport in samples with aspect ratio $L/W \gtrsim 1$ (in our case $L/W \approx 1.67$). In the opposite limit of wide and short samples, one expects minimum conductivity (rather than conductance), $\sigma_{xx,ABC}^{min} = \frac{8}{3\sqrt{3}}\frac{e^2}{h}$.

%
\begin{figure}[!t]
\vspace{-0.25in}
\includegraphics[width=1.\linewidth]{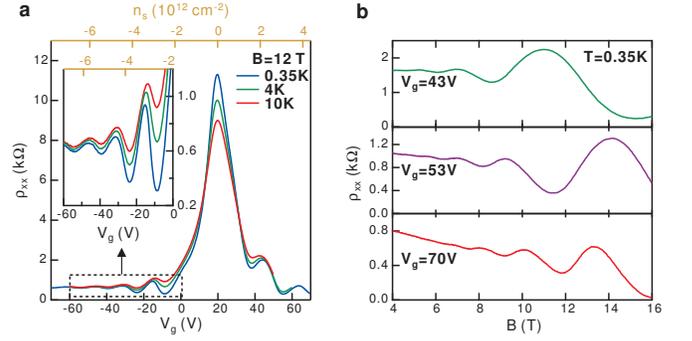}
\caption{Quantum Shubnikov-de Haas Oscillations in ABC trilayer. a, Longitudinal resistivity, $\rho_{xx}(V_g)$, for $B = 12$ T at 0.35 K, 4 K, and 10 K, top scale shows carrier density, $n_s(V_g)$. The inset is a zoom-in of the part surrounded by the dashed line. b, SdHO as a function of magnetic field for different $n_s(V_g)$. The decay of the SdHO magnitude with temperature and magnetic field is governed by the cyclotron mass of charge carriers, $m_c(n_s)$, and by the quantum scattering time, $\tau_{q}$.
 }
\label{Fig2:SdHO}
\vspace{-0.25in}
\end{figure}
%

The observed monolayer-like ``normal'' gate voltage dependence of the conductivity can be reconciled with the ``abnormal'' cubic dispersion of $l = 3$ chiral fermions if one accounts for super-screening of impurity Coulomb potential by such quasiparticles. Indeed, in the random phase approximation (RPA), the screened potential is governed by the diverging near $\varepsilon = 0$ density of states, rather than by the bare Coulomb interaction. As a result, the divergence in transport scattering cross-section is canceled by the vanishing screened potential, and one obtains charge-density-independent mobility in the ABC trilayer, $\mu_3 = 8/(\pi^2 N_i)$. Here $N_i$ is the concentration of Coulomb impurities, which are mainly located in the SiO$_2$ substrate. In monolayer graphene screening is much weaker, and simply amounts to multiplying the Coulomb impurity potential by a number, so it becomes $U(r) = -\hbar v \alpha/r$. If determined by the dielectric constant, $\Sigma \sim 4 - 10$, of glass substrate alone, $\alpha = e^2/(\hbar v\Sigma) \sim 0.5 - 0.2$. The mobility of the monolayer is $\mu_1 = 1/(\alpha^2 \pi^2 N_i)$. Hence, for an equal number of impurities, $\mu_3/\mu_1 = 8 \alpha^2$, and the mobility of $l = 3$ chiral fermions of the ABC trilayer could appear markedly lower than for massless fermions of the monolayer, if $\alpha$ is small.

%
\begin{figure}[!t]
\vspace{-0.25in}
\includegraphics[width=.8\linewidth]{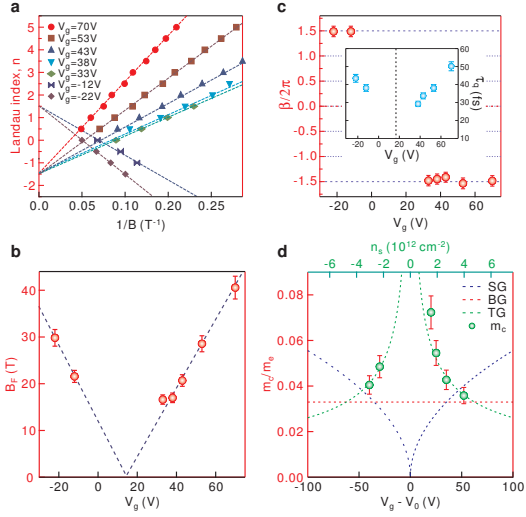}
\caption{ a, Landau fan diagram of SdHO. Points for integer $n$ correspond to $n$-th minimum, for half-integers, $n-1/2$, to $n$-th maximum of $\rho_{xx}(B)$ \cite{Novoselov2005,Zhang2005}. b, $B_F(V_g)$ dependence, the error bars show standard deviation, the dashed line is linear fit. c, The Berry phase of magneto-oscillations from fits in Fig. 3a. The inset shows quantum scattering time obtained by fitting the SdHO by the ALK formula. d, The cyclotron mass, $m_c$, of charge carriers as a function of $n_s(V_g)$, obtained from the same SdHO-ALK fits. Lines show the semi-classical result for linear (MG), quadratic (BG) and cubic (TG) spectra, with $\gamma_1 = 0.5$ eV obtained by fitting the measured $m_c(n_s)$.
 }
\label{Fig3:BerryPhase}
\vspace{-0.25in}
\end{figure}
%

Having thus established high quality of our ABC trilayer sample, we proceeded with magnetotransport measurements in a perpendicular field $B$. Figure 2 shows examples of SdH quantum magneto-oscillations at temperatures 0.35 K, 4 K and 10 K as a function of gate voltage for $B = 12$ T, and as a function of magnetic field for several $V_g$. In the semi-classical limit of small oscillations, they can be described by the Ando-Lifshitz-Kosevich (ALK) formula \cite{Coleridge1989},  $\frac{\Delta \rho_{xx}}{2\rho_0} = \frac{2\chi}{\sinh\chi} e^{-\frac{\pi}{\omega_c \tau_q}} \cos\left(\frac{2\pi B_F}{B} -\pi + \beta \right)$. Here, $\chi = 2\pi^2 k_BT/(\hbar \omega_c)$,  $k_B$ is Boltzmann's constant, $\tau_q$, is the quantum scattering time, $\omega_c = \frac{eB}{m_c}$ is the cyclotron angular velocity, $m_c=\frac{1}{2\pi}\frac{\partial S(\varepsilon)}{\partial \varepsilon} = m_{ABC}$ the cyclotron mass, $S(\varepsilon) = \pi p(\varepsilon)^2$ is the area in the momentum space of the orbit at the Fermi energy, $B_F = n_s\Phi_0/g_{LL}$, where $\Phi_0 = h/e \approx 4.14 \cdot 10^{-11}$ T·cm$^2$ is flux quantum and $g_{LL}$ is the LL degeneracy, and $\beta = l\pi$ is the Berry phase of the quasiparticles. Fig. 2 already reveals a clear hallmark of the large cyclotron mass and small scattering time, $\tau_q$, of $l = 3$ quasiparticles in the ABC trilayer. The decay of the magnitude of SdHO with increasing temperature, or decreasing magnetic field, or increasing charge density $n_s$, is strikingly fast, much faster than observed in monolayer or bilayer \cite{Novoselov2005,Zhang2005,Novoselov2006} (at least half a dozen oscillations are seen in similar measurements for the latter systems, while we can only reliably identify 3-4 in Fig. 2).

%
\begin{figure}[!t]
\includegraphics[width=.8\linewidth]{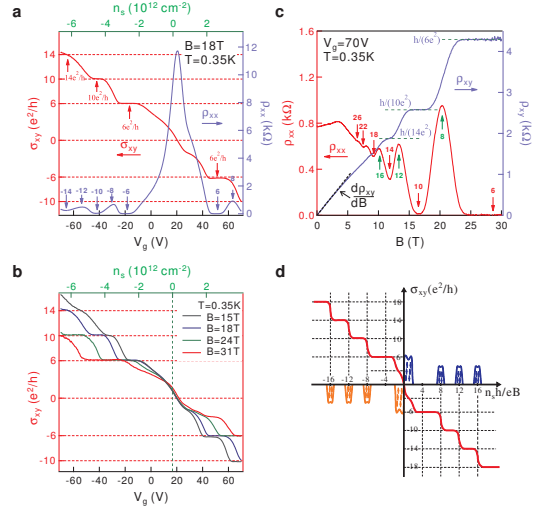}
\caption{QHE and Landau Level spectrum in ABC trilayer at T = 0.35 K. a, Longitudinal resistivity, $\rho_{xx}$ (right scale), and Hall conductivity, $\sigma_{xy}=\rho_{xy}/(\rho_{xx}^2+\rho_{xy}^2)$ (left scale) at $B = 18$ T. Vertical arrows with numbers show LL filling, $\nu  = n_sh/(eB)$, for the corresponding Quantum Hall states. b, $\sigma_{xy}(V_g)$ for several magnetic fields, $B$. c, $\rho_{xx}(B)$ (left scale), and $\rho_{xy}(B)$ (right scale) for $V_g = 70$ V. Quantized plateaus of $\rho_{xy}$ and zero $\rho_{xx}$ are clearly observed for $\nu = 6, 10$. Dashed horizontal lines show $h/(\nu e^2)$. Vertical arrows with numbers show LL fillings which can be identified in SdHO. d, Schematic illustration of the LL DOS in ABC stacked trilayer graphene.
 }
\label{Fig4:QHE}
\vspace{-0.25in}
\end{figure}
%

The simplest analysis of SdH oscillations is achieved by plotting positions of $\rho_{xx}$ minima and maxima as a function of $1/B$. The Landau fan diagram thus obtained is shown in Fig. 3a. It is already clear from linear fits in the figure that charge carriers in our device are characterized by a Berry phase of $3\pi$. Fit parameters are quantified in Fig. 3b,c, which show $B_F$ and $\beta$ as a function of $V_g$. Linear fits of $B_F(n_s)$ in Fig. 3b and using $dn_s/dV_g = 7.59 \cdot 10^{10}$ cm$^{-2}$ determined from the measured low-field Hall constant yield LL degeneracy $g_{LL} = 4 \pm 0.1$, in perfect agreement with the expected valley$\times$spin degeneracy. Similarly, the average Berry phase in Fig. 3c is $\beta = 2\pi \cdot (1.5 ± 0.1)$.

We can pursue an analysis of the measured SdHO data further by subtracting the smooth component of magnetoresistance, which is mainly governed by weak localization, and fitting the oscillating part with the ALK formula \cite{Supplementary}. Varying only 3 parameters, $B_F$,  $\tau_q$, and $m_c$, we obtain high fidelity fits by imposing a physical constraint, $\tau_q < 200$ fs. The obtained $B_F$ values are within the symbol size of those shown in Fig. 3c. The cyclotron mass of charge carriers obtained in such fitting is shown in Fig. 3d. It shows marked variation with carrier density, increasing towards the CNP, $n_s = 0$, which is opposite to that observed in monolayer graphene \cite{Novoselov2005,Zhang2005}. A fit to the divergent behaviour, $m_c = \frac{\gamma_1^2}{6\hbar v^3}(\pi n_s)^{-1/2}$ , expected for $l = 3$ quasiparticles in the ABC trilayer, is shown by the dotted line in Fig. 3d. It yields $\gamma_1 \approx (0.5 ± 0.1)$ eV (using $v \approx 10^6$ m/s), in very good agreement with the $\gamma_1$ value expected for few-layer graphene \cite{Ohta2007}. This agreement provides an important credibility check for our analysis, and for $\tau_q$ shown in the inset of Fig. 3c. In fact, it is quite remarkable that we are able to obtain the interlayer hopping parameter from the SdHO data.

Further support for the existence of $l = 3$ charge carriers is provided by the unusual QHE, which develops in our sample at T $\approx$ 0.35 K in fields above $\approx$ 15 T, Figure 4. Plateaux of Hall conductivity are observed at $\sigma_{xy} = \pm 6e^2/h, \pm 10e^2/h, ...$, with a step of $12e^2/h$ between the hole and electron gases across the N = 0 LL, confirming its 12-fold degeneracy. This, perhaps, is most clearly seen in Fig 4c, which shows the magnetic field dependence of the longitudinal, $\rho_{xx}$, and Hall, $\rho_{xy}$, resistivities, for $V_g = 70$ V ($n_s \approx 4 \cdot 10^{12}$ cm$^{-2}$). The minima of $\rho_{xx}$ occur near the center of $\rho_{xy}$ ($\sigma_{xy}$) plateaux, at LL filling factors $\nu = n_s \Phi_0/B = 6, 10, 14, ...$, as indicated by arrows. The behaviour near the N = 0 LL is in stark contrast to that in the monolayer and bilayer graphene, where the first QHE plateau develops at $\sigma_{xy} = \pm 2e^2/h$ ($\nu = 2$) and $\sigma_{xy} = \pm 4e^2/h$ ($\nu = 4$), respectively. There is only a weak anomaly in our sample for $n_s$ below the fist QHE plateau at $\nu = 6$, which can be associated with $\sigma_{xy} = \pm 3e^2/h$, $\nu = 3$, and is probably an indication of the developing spin-splitting, Fig. 4d.


To conclude, we have discovered that new, $l = 3$ chiral fermions indeed exist in the ABC trilayer graphene and govern properties of realistic graphene devices, so they can be detected in experiment. These quasiparticles accumulate a Berry phase of $3\pi$ along cyclotron trajectories and acquire unusual LL quantization in magnetic field, $\varepsilon^{\pm} = \pm \frac{\left( v \hbar \sqrt{2eB /(hc)} \right)^3}{\gamma_1^2} \sqrt{n(n-1)(n-2)}$, $n$ integer \cite{CastroNeto2009,Guinea2006,Aoki2007}, and therefore are revealed in magnetotransport measurements, such as we presented here. Not only our results provide experimental validation for the large body of important recent theoretical work \cite{CastroNeto2009,Guinea2006,Aoki2007,Ezawa2007,KoshinoAndo2007,KoshinoMcCann2009,Koshino2010,ZhangMacDonald2010}, but, perhaps more importantly, they greatly extend the perceived experimental limits, uncovering new exciting possibilities for future studies. Our findings are also significant because they establish the experimental feasibility of deploying the unusual properties of the $l = 3$ chiral fermions predicted theoretically \cite{Koshino2010}, such as super-screening and band gap tuning, in realistic, large-area graphene devices.

\begin{acknowledgments}

We thank T. Valla and E. Mendez for discussions and valuable comments. Material preparation and device processing was done at Brookhaven National Laboratory's Center for Functional Nanomaterials. This work was supported by the US DOE under Contract DE-\-AC02-\-98CH10886. The work of Yan Zhang is supported by NSF DMR-0705131. Magnetic field experiments were carried out at NHMFL, which is supported by NSF through DMR-0084173 and by the State of Florida.

\end{acknowledgments}


\begin{thebibliography}{10}


\bibitem{Novoselov2005}
K.~S. Novoselov, {\it et al.} 
 Nature \textbf{438}, 197 (2005).

\bibitem{Zhang2005}
Y. Zhang, {\it et al.} 
 Nature \textbf{438}, 201 (2005).

\bibitem{Novoselov2006}
K.~S.~Novoselov, {\it et al.}, 
Nat. Phys. \textbf{2}, 177-180 (2006).

\bibitem{CastroNeto2009}
A.~H. Castro~Neto, {\it et al.} 
 Rev. Mod. Phys. \textbf{81}, 109 (2009).

\bibitem{Guinea2006}
F.~Guinea, A.~H.~C.~Neto, N.~M.~R.~Peres,
Phys. Rev. B \textbf{73}, 245426 (2006);
Solid State Comm. \textbf{143}, 116, (2007).  

\bibitem{Aoki2007}
M.~Aoki, H.~Amawashi,
Solid State Comm. \textbf{142}, 123 (2007).   

\bibitem{Ezawa2007}
M.~Ezawa,
J. Phys. Soc. Jpn. \textbf{76}, 094701 (2007).

\bibitem{KoshinoAndo2007}
M.~Koshino, T.~Ando,
Phys. Rev. B \textbf{76}, 085425 (2007).

\bibitem{KoshinoMcCann2009}
M.~Koshino, E.~McCann,
Phys. Rev. B \textbf{80}, 165409 (2009).

\bibitem{Koshino2010}
M.~Koshino,
Phys. Rev. B \textbf{81}, 125304 (2010).

\bibitem{ZhangMacDonald2010}
F.~Zhang, S.~Bhagawan, H.~Min, A.~H.~MacDonald,
Phys. Rev. B \textbf{82}, 035409 (2010).

\bibitem{Craciun2009}
M.~F.~Craciun, {\it et al.}, 
Nat. Nanotech. \textbf{4}, 383 (2009).   

\bibitem{Zhu2009}
W.~Zhu, V.~Perebeinos, M.~Freitag, P.~Avouris,
Phys. Rev. B \textbf{80}, 235402 (2009);
%
Nano Letters \textbf{10}, 3572 (2010).  

\bibitem{Liu2010}
Y.~Liu, S.~Goolaup, C.~ Murapaka, W.~S.~Lew, S.~K.~Wong,
ACS Nano \textbf{4}, 7087 (2010).    

\bibitem{Bao2010}
W.~Bao, {\it et al.}, 
Phys. Rev. Lett. \textbf{105}, 246601 (2010).

\bibitem{Mak2010}
K.~F.~Mak, J.~Shan, T.~F.~Heinz,
Phys. Rev. Lett. \textbf{104}, 176404 (2010).

\bibitem{Friese1962}
E. J. Freise,
Nature \textbf{193}, 671 (1962).    

\bibitem{Zhang2009}
Y.~Zhang, {\it et al.}, 
Nature \textbf{459}, 820 (2009).    

\bibitem{low_n_note}
At very low energies, the cubic dispersion is modified by further hoppings, but estimates \cite{ZhangMacDonald2010} show that it holds for experimentally relevant densities, $n_s \gtrsim 10^{11}$ cm$^{-2}$.

\bibitem{Novoselov2004}
K.~S. Novoselov, {\it et al.}, 
 Science textbf{22}, 666 (2004).

\bibitem{Ferrari2006}
A.~C.~Ferrari, {\it et al.}, 
Phys. Rev. Lett. \textbf{97}, 187401 (2006).

\bibitem{Graf2007}
D. Graf, {\it et al.}, 
Nano Letters \textbf{7}, 238 (2007).   

\bibitem{Hao2010}
Y. F. Hao, {\it et al.},
Small \textbf{6}, 195 (2010).   

\bibitem{Lui2011}
Chun Hung Lui, {\it et al.}, 
Nano Letters \textbf{11}, 164 (2011).    

\bibitem{Coleridge1989}
P.~T.~Coleridge, R. Stoner, R. Fletcher,
Phys. Rev. B \textbf{39}, 1120 (1989);
%
P.~T.~Coleridge,
\emph{ibid} \textbf{44}, 3793 (1991).

\bibitem{Ohta2007}
T.~ Ohta, {\it et al.}, 
Phys. Rev. Lett. \textbf{98}, 206802 (2007).

\bibitem{Supplementary}
See Supplementary Material.




\end{thebibliography}

\end{document}